\documentclass[aps,prx,twocolumn,superscriptaddress,longbibliography]{revtex4-2}
\usepackage{amsmath,amssymb,bm,graphicx,booktabs,array}
\usepackage{xcolor}
\usepackage[colorlinks=true,allcolors=blue]{hyperref}
\usepackage{braket}
\usepackage{mathtools}
\usepackage{siunitx}
\usepackage[shortlabels]{enumitem}
\newcommand{\F}{\mathbb F_2}
\newcommand{\ketbra}[2]{|#1\rangle\!\langle#2|}

\newcommand{\im}{\operatorname{im}}

\newcommand{\simname}{\texttt{magna}}

\definecolor{revision}{rgb}{0.50,0.0,0.62}

\begin{document}
\title{Magna: simulating magic state factories as cheaply as Clifford circuits}

\author{Yifei Wang}
\affiliation{Chen-Ning Yang Institute for Advanced Study, Tsinghua University, Beijing, 100084, China}
\affiliation{Qosmos Technology (Beijing) Co., Ltd, Beijing 100176, China}
\affiliation{Beijing Key Laboratory of Cold Atom Quantum Computation, Tsinghua University, Beijing, 100084, China}

\author{Yingfei Gu}
\email[E-mail: ]{guyingfei@tsinghua.edu.cn}
\affiliation{Chen-Ning Yang Institute for Advanced Study, Tsinghua University, Beijing, 100084, China}
\affiliation{Beijing Key Laboratory of Cold Atom Quantum Computation, Tsinghua University, Beijing, 100084, China}

\date{October 15, 2026}

\begin{abstract}
Classical simulation of magic state factories, 
including distillation, cultivation, and code switching, 
is challenging because of their substantial non-Clifford content. 
We introduce \texttt{magna} (magic-state-factory assessment via global network algebra), a classical simulator 
that simulates such protocols as cheaply as Clifford circuits. 
\texttt{magna} expresses the figures of merit of a protocol, 
namely the acceptance rate and the fidelity of the accepted states, 
as tensor networks, which it reduces globally by exploiting 
a structural correspondence with transversal gates on quantum codes. 
\texttt{magna} further supports feedforward conditioned on 
arbitrary functions of mid-circuit measurement outcomes, 
including decoder outputs. 
We simulate Bravyi--Haah distillation with up to 100 output magic states 
at a cost that does not grow exponentially in the number of outputs, 
and achieve state-of-the-art sampling times for distance-5 magic state cultivation 
on both color codes and RP$^2$ surface codes. 
Finally, we validate nonlinear feedforward on a logical-level distillation protocol 
with 15 rounds of decoder-dependent feedforward.
\texttt{magna} aims to contribute to the key infrastructure 
for the large-scale evaluation and optimization of magic state factories.
\end{abstract}
\maketitle

\section{Introduction}\label{sec:intro}

Fault-tolerant quantum computation relies on quantum error-correcting codes, 
on which Clifford operations can typically be implemented with low overhead. 
By the Gottesman--Knill theorem, however, Clifford circuits acting on stabilizer states 
admit efficient classical simulation \cite{gottesmanHeisenbergRepresentationQuantum1998}, 
and universality therefore requires at least one non-Clifford operation. 
Such operations are commonly realized through gate teleportation 
using magic states as a consumable resource \cite{gottesmanDemonstratingViabilityUniversal1999,bravyiUniversalQuantumComputation2005}. 
Since the preparation of high-fidelity magic states constitutes a dominant contribution 
to the space-time overhead of fault-tolerant architectures, 
it is generally delegated to dedicated subroutines referred to as magic state factories. 
A variety of such protocols have been proposed, 
including magic state distillation \cite{bravyiUniversalQuantumComputation2005}, 
magic state cultivation \cite{gidneyMagicStateCultivation2024}, 
and code switching, which can be implemented by gauging \cite{andersonFaultTolerantConversionSteane2014,Bombín_2016}, 
transversal gates \cite{daguerreCodeSwitchingRevisited2025,wangEfficientFaulttolerantImplementations2024}, 
or code surgery \cite{wangNonCliffordQuantumGate2026}. 
Assessing and comparing these protocols requires accurate estimates of their acceptance rates 
and output fidelities under circuit-level noise, which are typically obtained through classical simulation. 
The non-Clifford content that renders these circuits computationally useful, 
however, is precisely what makes their classical simulation challenging.

Several classical simulators for noisy non-Clifford circuits have recently been developed, 
and they broadly fall into two classes. 
The first class performs state evolution, 
seeking compact representations of the intermediate states 
\cite{liSOFTHighperformanceSimulator2025,chaseClifftFastExact2026,fangSymFTUniversalFaultTolerant2026,hartwegComparingMagicState2026}. 
The cost of these methods is governed by the complexity of the intermediate states, 
which may carry substantial non-Clifford content 
even though they are irrelevant to the desired output of the protocol. 
The second class applies global simplifications to the entire circuit prior to evaluation 
\cite{kissingerSimulatingQuantumCircuits2021,pashayanFastEstimationOutcome2022,haenelTsimFastUniversal2026,surtiEfficientSimulationLogical2025}. 
These methods nevertheless remain exponentially costly in the number of output magic states, 
since the stabilizer rank of $\ket{T}^{\otimes k}$ is conjectured to grow exponentially in $k$ 
\cite{bravyiTradingClassicalQuantum2016}. 
Consequently, neither approach scales favorably to factories that produce many magic states in parallel.

In this work, we introduce \simname\ 
(magic-state-factory assessment via global network algebra)~\cite{magna}, a classical simulator tailored to magic state preparation protocols. 
\simname\ simulates the full non-Clifford circuits of these protocols 
without resorting to Clifford proxies, 
avoids an exponential cost in the number of output magic states, 
and supports causal sampling of nonlinear feedforward, 
a capability that is essential for fault-tolerant protocols 
but lacking in most existing simulators. 
We demonstrate these capabilities on a range of protocols (Sec.~\ref{sec:examples}), 
including magic state distillation \cite{bravyiMagicStateDistillation2012}, 
cultivation \cite{gidneyMagicStateCultivation2024,chenEfficientMagicState2025}, 
and code switching \cite{daguerreCodeSwitchingRevisited2025,wangNonCliffordQuantumGate2026}. 
For the Bravyi--Haah $(3k+8)$-to-$k$ distillation family \cite{bravyiMagicStateDistillation2012}, 
we simulate instances producing up to $k=100$ output magic states 
from 308 input $T$ states on 716 physical qubits, 
with neither the preparation time nor the per-draw cost growing exponentially in $k$. 
For magic state cultivation, we achieve state-of-the-art sampling times 
for the distance-5 protocols on both the color code (53 $T$ gates on 42 qubits) 
and the RP$^2$ surface code (101 $T$ gates on 61 qubits). 
Finally, we validate the treatment of nonlinear feedforward 
on a logical-level magic state distillation protocol based on the surface code, 
which involves 15 rounds of decoder-dependent feedforward, 
and recover the exact theoretical logical error rates under a simple noise model.

Our approach rests on two key observations. 
First, since the acceptance rate and the output fidelity are the only quantities of interest, 
we compute these scalars directly by expressing each as a tensor network, 
so that neither the intermediate states nor the output state need ever be constructed. 
Second, we reduce these tensor networks globally, 
exploiting a structural correspondence with transversal non-Clifford gates on quantum codes. 
This reduction identifies a basis in which the non-Clifford resource 
is concentrated on only a few degrees of freedom. 
We further show that the reduced structure is preserved under Pauli insertions, 
which allows a single reduction to be reused across all sampled error realizations 
and makes nonlinear feedforward tractable.

\section{Simulation framework}\label{sec:algorithm}

\subsection{General workflow}\label{sec:workflow}

A magic state preparation protocol aims to produce one or more magic states
with high fidelity, potentially requiring postselection.
Such a protocol is characterized by
its acceptance rate (equal to one if no postselection is required)
and the fidelity of the accepted output with the ideal target state $\ket{\psi_{\text{tar}}}$,
which are defined as follows.

Let $\rho$ describe the state produced by the noisy protocol before postselection,
with auxiliary registers representing measurement records and discarded qubits.
The acceptance rule is a binary function $\mathrm{ps}(m)$ of the reported measurement outcomes $m$.
The projector $\Pi_{\text{ps}}$ selects records satisfying $\mathrm{ps}(m)=\textsc{accept}$.
Without postselection, we take $\Pi_{\text{ps}}=I$.
The quantities of interest are
\begin{equation}
      p(\text{acc})=\operatorname{tr}(\rho\Pi_{\text{ps}}),\quad
      F=p(\text{acc})^{-1}\operatorname{tr}\braket{\psi_{\text{tar}}|\Pi_{\text{ps}}\rho\Pi_{\text{ps}}|\psi_{\text{tar}}}.
      \label{eq:required-output}
\end{equation}
For the fidelity $F$, the overlap is taken on the output register,
and the trace is taken over the auxiliary registers.
We also refer to the infidelity $1-F$ as the logical error rate.

Under stochastic Pauli noise, we estimate these quantities by sampling error realizations.
For a realization $e$ drawn with probability $p(e)$,
let $\ket{\psi_e}$ be the state of the output and all auxiliary registers,
therefore $\rho=\sum_e p(e)\ketbra{\psi_e}{\psi_e}$ and 
\begin{align}
      p(\text{acc})&=\sum_e p(e)\underbrace{\braket{\psi_e|\Pi_{\text{ps}}|\psi_e}}_{:=p({\rm acc}\mid e)},
      \label{eq:acc-from-sample}\\
      F&=p(\text{acc})^{-1}\sum_e p(e)
      \underbrace{
      \bigl\|\bra{\psi_{\text{tar}}}\Pi_{\text{ps}}\ket{\psi_e}\bigr\|^2}_{:=F({\rm acc},e)}.
      \label{eq:F-from-sample}
\end{align}
Our goal is to efficiently evaluate the acceptance probability $p(\text{acc}\mid e)$ and the target overlap $F({\rm acc},e)$, where 
the squared norm sums over the remaining auxiliary registers.

Now, we evaluate the expressions for $p(\text{acc})$ and $F$ through a global method that transforms the circuits to networks, i.e. for each $e$, we construct an acceptance network $\mathcal N_{\text{acc}}(e)$ and a fidelity network $\mathcal N_{\text{fid}}(e)$ for $p(\text{acc}\mid e)$ and $F({\rm acc},e)$ respectively. One key trick is that we will generate these networks based on a pair of \emph{reference networks} $\mathcal N_{\text{acc}}$, $\mathcal N_{\text{fid}}$ for noiseless circuits, in an incremental manner. 
In other words, we will proceed in the following way
\begin{enumerate}
\item {\it Preparation stage}: 
construct $\mathcal N_{\text{acc}}$ and $\mathcal N_{\text{fid}}$ for the noiseless circuit. This construction involves strategic global simplification, utilizing the quantum error-correction structure of the magic state preparation circuit as mentioned in the introduction. 
This part will be explained in Sec.~\ref{sec:reference-networks}. 
This stage has polynomial cost in the circuit size.

\item {\it Update the network with Pauli insertions}:
now for each sampled error realization $e$, we obtain the networks $\mathcal N_{\text{acc}}(e)$ and $\mathcal N_{\text{fid}}(e)$ by inserting the Pauli errors into the reference networks.
The key is that we can show that Pauli insertions preserve the structures needed to reuse the construction, ensuring that the complexity of the full simulation is controlled by the reference networks. We refer to this phenomenon as the Pauli-covariance of the network, detailed in Sec.~\ref{sec:pauli-stability};

\item {\it Feedforward}: 
how feedforward enters this workflow depends on its classical rule,
the function $f$ of previous measurement outcomes that determines whether a gate is applied.
We focus on cases in which the applied gates are Pauli or Clifford.
If the function $f$ is linear (allowing offsets),
we represent the feedforward as a controlled quantum gate and incorporate it into the networks $\mathcal{N}_{\text{acc}}$ and $\mathcal{N}_{\text{fid}}$.
For nonlinear $f$, such as the decoder used in logical-level magic state distillation,
it is
in general 
difficult to rewrite the control logic as quantum gates,
and such a representation can introduce many non-Clifford resources that resist global reduction.
We instead sample the required reported outcomes and evaluate the classical rule once its inputs are available.
This demands a third reference network, the sampling network $\mathcal{N}_{\text{samp}}$ (Fig.~\ref{fig:sampling-network}),
for generating the desired measurement outcomes.
As shown in Sec.~\ref{sec:nonlinear-feedforward}, all three networks can still be updated incrementally
across different error realizations and mid-circuit measurement outcomes.
\end{enumerate}

\begin{figure}[t]
\centering
\includegraphics{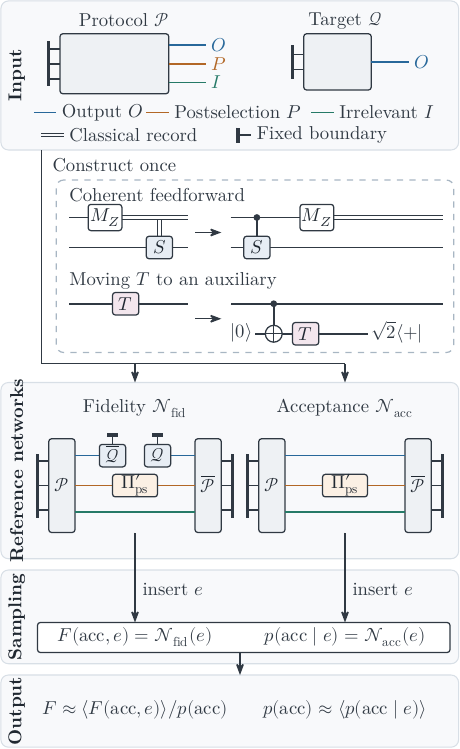}
\caption{Simulation workflow.
The protocol circuit $\mathcal P$ and the target preparation circuit $\mathcal Q$ are written into tensors with fixed beginnings (bars)
and output ($O$), postselection ($P$), and irrelevant ($I$) ending legs.
We construct the reference networks $\mathcal{N}_{\text{acc}}$ and $\mathcal{N}_{\text{fid}}$ in the preparation stage,
which join $\mathcal P$ to its conjugate $\overline{\mathcal P}$,
project the postselection legs with $\Pi_{\text{ps}}'$, and trace the irrelevant legs.
The fidelity network attaches the target to the output legs, which the acceptance network traces.
The inset shows two rewriting gadgets in the network construction:
a Clifford correction controlled by a measurement outcome becomes a coherent control applied before the measurement (top),
and a $T$ gate moves onto a fresh auxiliary qubit (bottom).
For each draw, a fault $e$ is inserted into both networks which updates them incrementally
to obtain $\mathcal{N}_{\text{acc}}(e)$ and $\mathcal{N}_{\text{fid}}(e)$,
whose contraction gives $p(\text{acc}\mid e)$ and $F(\text{acc},e)$.
Averaging over all draws produces the output.}
\label{fig:reference-compilation}
\end{figure}

\subsection{Initialization of reference networks}\label{sec:reference-networks}

Protocols without nonlinear feedforward rules only need two reference networks
$\mathcal{N}_{\text{acc}}$ and $\mathcal{N}_{\text{fid}}$.
Figure~\ref{fig:reference-compilation} shows these networks and, in its inset, the rewriting gadgets used to construct them.
Our algorithm is applicable to circuits built from the following components.
\begin{enumerate}
      \item Initialization in computational-basis states;
      \item Clifford gates and gates from the third level of the Clifford hierarchy (CH): $T$, $\mathrm{C}S$, and $\mathrm{CC}Z$;
      \item Computational-basis measurements;
      \item Reset or discard of qubits;
      \item Clifford or Pauli feedforward controlled by a linear function of measurement outcomes, including mid-circuit outcomes;
      \item Postselection on a linear function of measurement outcomes.
\end{enumerate}
The target state is also specified by a virtual preparation circuit built from the same components.
This circuit only defines the ideal state and carries no fault-tolerance requirement.
The noiseless protocol circuit itself is one valid choice,
but a much smaller circuit preparing the same state
improves the simulation efficiency by reducing the size of the network.

The construction starts by reorganizing the protocol circuit in three steps.
\begin{enumerate}
      \item We replace each reset by discarding the old qubit and initializing a fresh one.
      \item We replace each classical feedforward control by a coherent control,
      postponing the measurement until after the controlled gate
      [Fig.~\ref{fig:reference-compilation}, inset, top].
      A Clifford correction thereby becomes a non-Clifford controlled gate, such as controlled-$S$,
      but the whole protocol becomes a single coherent circuit,
      and the added non-Clifford resources enter the global reduction below.
      \item We apply a virtual Clifford $C_{\text{rec}}$ to the measurement records
      so that each independent postselection condition fixes the value of a single record leg.
      Because the conditions are linear in the outcomes, $C_{\text{rec}}$ can be chosen as a network of CNOT gates.
      The acceptance condition then acts through $\Pi_{\text{ps}}'=C_{\text{rec}}\Pi_{\text{ps}}C_{\text{rec}}^\dagger$,
      a product of single-leg projectors onto the accepted values.
\end{enumerate}
We regard all fixed preparations as the ``beginning legs''
and all measurements, discarded systems, and requested outputs as the ``ending legs.''
The ending legs fall into three classes:
\begin{enumerate}
      \item Output legs, which carry the requested output state;
      \item Postselection legs, whose specified outcomes determine acceptance;
      \item Irrelevant legs, whose individual values are no longer required by the requested event or any remaining classical processing.
\end{enumerate}
Irrelevant legs can arise from the records used for coherent feedforward,
from record legs that $\Pi_{\text{ps}}'$ leaves unfixed after the change of coordinates,
or from other redundant measurement outcomes.
They may remain correlated with the output.

Each reference network joins the reorganized protocol circuit to its conjugate
and closes each class of ending legs with a boundary condition,
as shown in Fig.~\ref{fig:reference-compilation}.
In the acceptance network $\mathcal{N}_{\text{acc}}$,
the irrelevant and output legs are connected between the two copies,
and the postselection legs are joined through $\Pi_{\text{ps}}'$.
The fully contracted network gives $p(\text{acc}\mid e)$ of Eq.~\eqref{eq:acc-from-sample} for the noiseless circuit.
The fidelity network $\mathcal{N}_{\text{fid}}$ differs only in its output legs,
which are connected to the output legs of the target circuit and its conjugate instead of to each other.
It gives $F(\text{acc},e)\braket{\tilde\psi_{\text{tar}}|\tilde\psi_{\text{tar}}}$,
where $\ket{\tilde\psi_{\text{tar}}}$ is the output of the virtual target circuit.
The normalization factor $\braket{\tilde\psi_{\text{tar}}|\tilde\psi_{\text{tar}}}$,
which can be nontrivial when postselection exists in the target circuit,
can be computed from a similar acceptance network for that circuit.

\subsection{Reduction via global network algebra}\label{sec:global-reduction}

We now reduce each reference network to a form 
whose evaluation cost is governed only by its genuine non-Clifford content. 
The reduction exploits the algebraic structure underlying transversal non-Clifford gates 
on quantum error-correcting codes, and proceeds in two stages:
\begin{enumerate}
    \item We push all Clifford operations to the boundaries of the network, 
    leaving a bulk of mutually commuting $\pi/8$ Pauli rotations that can be diagonalized simultaneously. 
    The stabilizer boundaries then confine the contributing computational-basis labels 
    to an affine subspace (linear subspace shifted by a vector), which plays the role of a code space. 
    \item We identify the directions within this subspace along which the diagonal phase is purely Clifford, 
    analogous to the $X$-type stabilizers of a CSS code, and absorb them into the boundaries, 
    so that the remaining non-Clifford dependence is confined to a few ``logical'' coordinates. 
\end{enumerate}
The result is an expansion of the network into a sum of stabilizer terms, 
whose number, the representation rank, sets the cost of evaluating the network. 
Both stages require only Gaussian elimination over $\F$ and stabilizer-state contractions, 
and therefore take time polynomial in the circuit size.

We begin with the first stage, 
which separates the Clifford part of each network, handled efficiently by stabilizer methods, 
from a layer of non-Clifford rotations. 
To this end, we express all unitary gates in terms of Pauli rotations 
$R_P(\theta)\coloneqq\exp(-\mathrm{i}\theta P)$. 
The non-Clifford gates permitted in the protocol circuit, namely $T$, $\mathrm{C}S$, and $\mathrm{CC}Z$, 
decompose into $\pi/8$ Pauli rotations and Clifford gates, 
as do the controlled Clifford gates introduced by coherent feedforward. 
For instance, the $T$ gate equals a single Pauli rotation up to a global phase, 
$T=\mathrm{e}^{\mathrm{i}\pi/8}R_Z(\pi/8)$. 
For any Clifford unitary $C$, 
\begin{equation}
 C R_P(\theta)=R_{CPC^\dagger}(\theta)\,C,
 \label{eq:clifford-transport}
\end{equation}
and since $CPC^\dagger$ is again a Pauli operator, 
commuting a Clifford gate past a rotation changes only the rotation's Pauli axis.

Applying Eq.~\eqref{eq:clifford-transport} repeatedly, 
we transport all Clifford gates to the beginning boundaries 
of both the protocol and target circuits, 
so that only $\pi/8$ rotations remain in the bulk of the networks. 
If the Pauli axes of these rotations do not all commute, 
we use the gadget shown at the bottom of the inset of Fig.~\ref{fig:reference-compilation} 
to move the noncommuting rotations onto separate auxiliary qubits. 
The resulting networks consist of stabilizer-state projectors, 
located at the beginning boundaries and at the internal gadget contractions, 
together with a bulk of mutually commuting $\pi/8$ rotations. 
This use of magic-state auxiliaries combined with global stabilizer contraction 
is closely related to the circuit compression employed in the outcome-probability algorithms 
of Ref.~\cite{pashayanFastEstimationOutcome2022}.

We next diagonalize the remaining bulk rotations. 
Since their Pauli axes mutually commute, 
there exists a Clifford unitary $D$ that maps all of them simultaneously to signed products of $Z$ operators, 
rendering the rotation layer diagonal in the computational basis. 
Writing the layer as $U=D^\dagger(DUD^\dagger)D$ 
and absorbing the outer Clifford factors $D$ and $D^\dagger$ into the boundary projectors, 
we obtain a diagonal bulk enclosed by stabilizer boundaries. 
These boundaries restrict the computational-basis labels $x\in\F^n$ that contribute to the network: 
a label contributes only if $\braket{x|\Pi|x}>0$ for every boundary projector $\Pi$. 
This restriction is equivalent to an inhomogeneous linear system on $x$,
whose solution set is either the empty set or an affine subspace of $b+V$, 
with $b\in\F^n$ and $V\subseteq\F^n$ a linear subspace. 
We refer to it as the \emph{affine support} of the rotation layer; 
the network need only be evaluated on this support.

In the second stage, we simplify the remaining rotation layer on its affine support. 
We characterize the diagonal gate by its phase polynomial $\varphi:\F^n\to\mathbb{Z}_8$, defined through
\begin{equation}
 DUD^\dagger\ket{x}=\gamma\,\omega^{\varphi(x)}\ket{x},
\end{equation}
where $\omega=\mathrm{e}^{\mathrm{i}\pi/4}$ and $\gamma$ is a global scalar independent of $x$. 
Because $U$ is a product of commuting $\pi/8$ Pauli rotations, 
$DUD^\dagger$ is a diagonal gate in the third level of the Clifford hierarchy, 
so that $\varphi(x)=\sum_i a_ix_i+2\sum_{i<j}b_{ij}x_ix_j+4\sum_{i<j<k}c_{ijk}x_ix_jx_k \bmod 8$ 
with integer coefficients $a_i$, $b_{ij}$, and $c_{ijk}$~\cite{cuiDiagonalGatesClifford2017}.

Our simplification of $\varphi$ is motivated by the structure of transversal gates on CSS codes. 
Consider a CSS code with check matrices $H_X$ and $H_Z$. 
Its code states are supported on the computational-basis labels in $\ker H_Z$, 
and its $X$-type stabilizers generate the subspace $\im H_X^{\mathsf T}\subseteq\ker H_Z$. 
If a diagonal physical gate implements a logical non-Clifford gate, such as $\overline{T}$, 
on the logical space $\ker H_Z/\im H_X^{\mathsf T}$, 
then its phase varies only in a Clifford manner along $\im H_X^{\mathsf T}$, 
and its non-Clifford action is confined to the logical degrees of freedom.

In our setting, the affine support plays the role of the code space, 
with $V$ in place of $\ker H_Z$, and the bulk acts on it as a diagonal gate. 
Our task is thus to identify the counterpart of $\im H_X^{\mathsf T}$, 
namely a subspace $K\subseteq V$ along which the phase varies only in a Clifford manner. 
To this end, we restrict the phase polynomial to the affine support, 
writing (with slight abuse of notation) $\varphi(y)\coloneqq\varphi(b\oplus y)$ for $y\in V$, 
and define the phase trilinear form $\sigma_\varphi:V^3\to\F$ through
\begin{equation}
 \Delta_u\Delta_v\Delta_w\varphi(y)=4\,\sigma_\varphi(u,v,w)\pmod 8,
 \label{eq:phase-trilinear}
\end{equation}
where $\Delta_u G(y)\coloneqq G(y\oplus u)-G(y)$ denotes the discrete derivative along $u\in V$. 
The form $\sigma_\varphi$ is a basis-independent version of the signature tensor 
of Ref.~\cite{heyfronEfficientQuantumCompiler2018}, 
and it vanishes identically if and only if the phase is Clifford 
\cite{cuiDiagonalGatesClifford2017,heyfronEfficientQuantumCompiler2018}. 
The Clifford subspace is therefore the radical of $\sigma_\varphi$,
\begin{equation}
 K=\{u\in V:\sigma_\varphi(u,v,w)=0\ \ \forall\, v,w\in V\}.
\end{equation}
Choosing coordinates on $V$ adapted to $K$ and to the quotient $W=V/K$, 
we absorb both the Clifford phase and the affine change of coordinates into the boundaries. 
The remaining phase then depends only on the ``logical'' coordinates in $W$, 
so that the network becomes a sum of $2^{d_W}$ stabilizer terms, where $d_W=\dim W$.

An alternative expansion can yield fewer terms. 
The remaining phase on $W$ can be generated by $t$ rotations of angle $\pi/8$, 
each of which can be implemented by consuming one $\ket{T}$ state \cite{litinskiGameSurfaceCodes2019}. 
The state $\ket{T}^{\otimes t}$ admits known decompositions into $c(t)$ stabilizer states, 
with $c(t)$ potentially much smaller than $2^t$ \cite{bravyiTradingClassicalQuantum2016}. 
Since $t$ may be larger or smaller than $d_W$, neither expansion is uniformly preferable. 
We therefore adopt whichever yields fewer terms, 
and define the \emph{representation rank} of the network as $\min\{2^{d_W},c(t)\}$.
Note that this rank is neither the trivial rank one of the network,
nor the stabilizer rank of the output state.
It measures the non-Clifford content that remains after our polynomial-time global reduction.

\subsection{Pauli-covariance of the networks}
\label{sec:pauli-stability}

We now study how the networks change under insertions of Pauli operators to the original circuits,
establishing the Pauli-covariance of the networks.
These insertions can be viewed as stochastic Pauli faults
or as corrections chosen by classical control functions.
From the first viewpoint, this subsection shows that the covariance makes sampling efficient
for protocols with linear feedforward and postselection,
with the resource bound certified once at construction of the network.
The second viewpoint will be used in the next subsection to handle nonlinear control rules.

We return to the reorganized protocol circuit,
before the global simplification.
We propagate the inserted Paulis through the circuit using two rules:
\begin{enumerate}
      \item Through a Clifford gate $C$, a Pauli $P$ becomes $Q=CPC^\dagger$,
      so $CP=QC$.
      \item Through a $\pi/8$ rotation, a Pauli changes the rotation sign according to
      $P R_Q(\pi/8)=R_Q((-1)^{\langle P,Q\rangle}\pi/8)P$,
      where $\langle P,Q\rangle$ is 1 when $P$ and $Q$ anticommute and 0 otherwise.
\end{enumerate}
The propagated operators flip the eigenvalues of some boundary stabilizers
and the signs of some bulk rotations, and they can contribute scalar phases, which we keep track of.
The Pauli operators defining the boundary stabilizers and the bulk rotation axes remain unchanged.

To understand how these changes affect the networks,
suppose we repeat the reference-network reduction on the circuit with Pauli insertions.
The flipped boundary stabilizer eigenvalues change only the offset of the affine label set of each boundary,
so the affine support changes from $b+V$ to $b_e+V$.
We can therefore reevaluate the supported phase polynomial in the same coordinates $y\in V$.
Since all rotations belong to one commuting layer, reversing the sign of a $\pi/8$ rotation
changes the layer only by a Clifford $\pi/4$ rotation about the same axis,
and shifting the affine offset also adds only a Clifford phase.
The updated phase thus has the form $\varphi_e(y)=\varphi(y)+E_e(y)$,
where $E_e$ is a Clifford phase.
By definition, the phase trilinear form is linear in $\varphi$ and vanishes on Clifford phases,
so $\sigma_{\varphi_e}=\sigma_\varphi$, which leaves both $K$ and $W=V/K$ unchanged.
The additional Clifford phase $E_e$ is then absorbed into the stabilizer boundaries,
which completes the reduction.

This analysis gives an incremental update strategy:
we propagate the inserted Paulis, update the affine offsets,
and absorb the resulting Clifford phase into the prepared boundaries.
The updated networks therefore require no more terms than the reference representations,
and all scalar weights are tracked exactly.
We call this \emph{Pauli-covariance of the networks}.
Evaluating each draw thus costs polynomial time in the circuit size per term,
and the number of terms is bounded by the representation rank, which is known as soon as the networks are constructed.

For stochastic Pauli noise with linear control rules,
we estimate the acceptance rate and the accepted fidelity $F$
using Eqs.~\eqref{eq:acc-from-sample} and~\eqref{eq:F-from-sample}.
To achieve additive error $\epsilon$ with failure probability at most $\delta$,
it suffices to take
\begin{equation}
      n_{\text{acc}}=O\!\left(\frac{\log(1/\delta)}{\epsilon^2}\right),\qquad
      n_F=O\!\left(\frac{\log(1/\delta)}{p(\text{acc})\epsilon^2}\right)
\end{equation}
independent fault draws for acceptance and fidelity, respectively, with $p(\text{acc})>0$.
These sample counts, combined with polynomial-time incremental updates and stabilizer contractions,
give a total cost polynomial in the circuit size, the representation ranks of the reference networks,
$p(\text{acc})^{-1}$, $\epsilon^{-1}$, and $\log(1/\delta)$.

\subsection{Nonlinear feedforward and postselection}
\label{sec:nonlinear-feedforward}

\begin{figure*}[tp]
\centering
\includegraphics{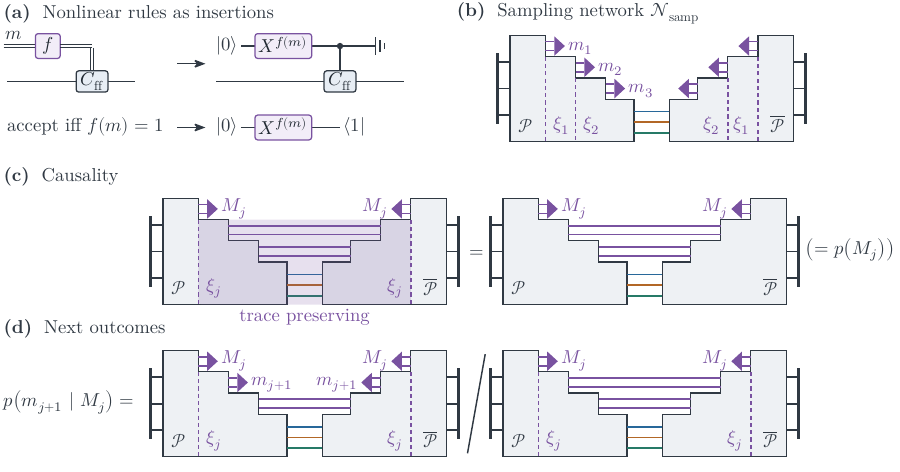}
\caption{Nonlinear rules and causal history sampling.
(a) A Clifford gate $C_{\text{ff}}$ controlled by a nonlinear function $f$ of earlier outcomes $m$
becomes a fixed controlled-$C_{\text{ff}}$ gate whose control is a virtual qubit,
which is prepared in $\ket0$, receives the Pauli insertion $X^{f(m)}$, and is then traced (top).
A nonlinear acceptance condition becomes the same insertion on a virtual qubit
that the acceptance and fidelity networks project onto $\ket1$ (bottom).
(b) The sampling network joins the rewritten protocol $\mathcal P$ to its conjugate $\overline{\mathcal P}$,
traces the output, postselection, and irrelevant legs (colors as in Fig.~\ref{fig:reference-compilation}),
and fixes all sampling legs to the outcomes $m_j$ (triangles).
Each half is drawn as a staircase whose successive steps emit the sampling legs of successive rounds,
and a dashed line marks where the feedforward $\xi_j$ determined by the first $j$ rounds acts.
(c) The first step collects the legs of $L_j$, which are fixed to the history $M_j$ (triangles),
and all later legs are traced.
The part after round $j$ (shaded) is then trace preserving,
so the value does not depend on $\xi_j$,
and equals $p(M_j)$ as proven in Eq.~\eqref{eq:history-network-law}.
(d) Fixing the outcomes $m_{j+1}$ of the next round as well gives their conditional distribution,
Eq.~\eqref{eq:history-conditional}.}
\label{fig:sampling-network}
\end{figure*}

We now treat feedforward and postselection controlled by a nonlinear function $f$
of mid-circuit measurement outcomes,
such as a decoder that proposes a correction or a logical decision from the measured syndromes.
As in Sec.~\ref{sec:reference-networks}, we require the controlled operations to be Pauli or Clifford.
Here, the main difficulty is that a global method simplifies the whole circuit before any outcome is known,
so such a rule must be represented coherently,
which, as noted in Sec.~\ref{sec:workflow}, can introduce many non-Clifford resources.
We resolve this by sampling the outcomes required by $f$ in causal order from a globally constructed network,
evaluating $f$ classically on the sampled values,
and returning its decisions to the network as updates.
Two questions must be addressed for this strategy to work:
how to incorporate nonlinear decisions into incremental reference-network updates,
and how to recover the physical probability of a measurement history
from a network whose feedforward assignments change during sampling.

For the first question, we express each update conditioned on a nonlinear signal $f(m)$
as a Pauli insertion, so that the Pauli-covariance introduced in Sec.~\ref{sec:pauli-stability} applies.
Pauli feedforward already has this form.
For Clifford feedforward $C_{\text{ff}}^{f(m)}$, we prepare a virtual qubit in $\ket0$,
apply $X^{f(m)}$ to it, use it to control a fixed controlled-$C_{\text{ff}}$ gate, and then trace it
[Fig.~\ref{fig:sampling-network}(a), top].
For nonlinear postselection accepting when $f(m)=1$,
we prepare another virtual qubit in $\ket0$, apply $X^{f(m)}$,
and project it onto $\ket1$ in the acceptance and fidelity networks
[Fig.~\ref{fig:sampling-network}(a), bottom].
We denote these conditional insertions collectively by $\xi$.
Apart from $\xi$, the circuit then consists only of the components of Sec.~\ref{sec:reference-networks},
including its linear feedforward and postselection,
and $\xi$ acts as an ordinary Pauli insertion once the outcomes it depends on are sampled.

It remains to sample these outcomes with their physical probabilities,
which is the second question.
For this purpose, we construct a third reference network $\mathcal N_{\text{samp}}$.
We call the measurement legs needed by these rules \emph{sampling legs}.
We join the rewritten protocol to its time-reversed copy along the output, irrelevant,
and postselection legs, including the virtual qubits of nonlinear postselection,
and initially set the insertions $\xi$ to zero [Fig.~\ref{fig:sampling-network}(b)].
All sampling legs are fixed to reference outcomes allowed by the stabilizer support of the network,
in the fidelity and acceptance networks as well.
A different measurement outcome then enters as Pauli $X$ insertions on the legs where it differs from the reference,
which, like faults, are transported by the Pauli-covariance of Sec.~\ref{sec:pauli-stability}.

We now show how $\mathcal N_{\text{samp}}$ gives the physical probability of the history.
We assume that faults are drawn independently of the measurement outcomes,
so that we can fix a fault draw $e$, include it in $\mathcal N_{\text{samp}}$ as Pauli insertions,
and suppress it from the notation.
All probabilities below are conditional on $e$.
We arrange the sampling legs $L$ into cumulative sets
$\varnothing=L_0\subset L_1\subset\cdots\subset L_T=L$ in causal order,
so that the feedforward determined by the outcomes on $L_j$ acts only after all legs of $L_j$ are measured.
Let $m_j$ denote the outcomes on $L_j\setminus L_{j-1}$,
$M_j=(m_1,\ldots,m_j)$ the history up to step $j$,
and $\xi_j$ the feedforward insertions determined by that history, with $\xi_0=0$.
For the derivation, we write $\mathcal N_{\text{samp}}[\xi_i](M_j)$ for the value of the sampling network
with the insertions $\xi_i$, those not yet determined set to zero,
the legs of $L_j$ fixed to $M_j$, and all other sampling legs traced.
For a complete history $M=M_T$, all sampling legs are fixed, as in the network constructed above.

These values are related to the physical probabilities by causality.
Feedforward applied after all legs of $L_j$ are measured acts on a part of the network whose legs are all traced.
Because the postselection legs are also traced rather than projected,
this part is trace preserving and does not change the distribution of the outcomes on $L_j$
[Fig.~\ref{fig:sampling-network}(c)].
Consequently, $\mathcal N_{\text{samp}}[\xi_i](M_j)$ does not depend on $i$ for $i\ge j-1$,
and $\mathcal N_{\text{samp}}[\xi_j]$ with $L_j$ fixed to $M_j$ gives the physical distribution
of the next outcomes $m_{j+1}$ up to normalization [Fig.~\ref{fig:sampling-network}(d)],
\begin{equation}
      p(m_{j+1}\mid M_j)=\frac{\mathcal N_{\text{samp}}[\xi_j](M_{j+1})}{\mathcal N_{\text{samp}}[\xi_j](M_j)}.
      \label{eq:history-conditional}
\end{equation}
Since the fully traced network is normalized, $\mathcal N_{\text{samp}}[\xi_0](M_0)=1$,
the product of these conditional probabilities telescopes to
\begin{equation}
      p(M_j)=\mathcal N_{\text{samp}}[\xi_j](M_j).
      \label{eq:history-network-law}
\end{equation}
In particular, the probability of a complete history $M=M_T$ is a single contraction
of the constructed network, with its sampling legs fixed to $M$ and the feedforward of $M$ inserted,
which later normalizes the acceptance and fidelity values of the history.

These relations allow us to sample complete histories.
We propose a complete history $M$ from a proposal distribution $q(M)$,
insert its feedforward, and correct by the weight $w=p(M)/q(M)$
obtained from Eq.~\eqref{eq:history-network-law}.
However, a poorly chosen $q$, such as the uniform distribution,
would frequently produce histories of zero probability,
which makes the sampling inefficient.
A simple improvement of $q$ is to choose a proposal that avoids
the histories excluded by the stabilizer support.
We call this the \emph{affine-support method} in our algorithm.
Given $M_{j-1}$ and $\xi_{j-1}$, the prepared stabilizer support data determine
an affine space that contains all possible next outcomes $m_j$.
We draw $m_j$ uniformly from this space,
so that $q(M)$ is the product of the inverse sizes of these spaces.
During sampling, we update only $\mathcal N_{\text{samp}}$,
and once the history is complete, we apply the accumulated insertions to the fidelity and acceptance networks.
Each draw costs time polynomial in the representation ranks of the reference networks and in the circuit size,
since the Pauli updates follow Sec.~\ref{sec:pauli-stability}
and the affine support requires only polynomial-time linear algebra.

The affine-support proposal does not exclude the possibility of a mismatch between $q$ and $p$ due to non-Clifford interference, 
which increases the variance of the weighted estimates.
When $\mathcal N_{\text{samp}}$ has a single stabilizer term ($d_W=0$), there is no interference,
so $q=p$ and every history has weight one,
which is the case of all examples in Table~\ref{tab:examples-overview}.
When the mismatch makes the sampling inefficient,
each round can instead be drawn from its exact conditional distribution, Eq.~\eqref{eq:history-conditional},
so that every history has weight one.
This \emph{exact-marginal method}, also implemented in \simname,
requires another sampling network that evaluates the partially traced values in Eq.~\eqref{eq:history-conditional},
which we do not expand here.

Finally, we return to the assumption that faults are independent of the measurement outcomes.
This assumption does not always hold in practice.
A fault attached to a feedforward gate occurs only when the gate is applied,
so its distribution depends on the history.
In the coherent representation of Sec.~\ref{sec:reference-networks}, such a fault becomes a controlled Pauli operator,
a Clifford insertion to which the covariance of Sec.~\ref{sec:pauli-stability} does not apply,
even when the control is linear.
\simname\ handles these gates with the sampling discussed in this section,
treating their control outcomes as sampling legs.
Once the sampled history determines whether a gate is applied,
its fault is drawn from the corresponding conditional distribution and inserted as a Pauli operator together with $\xi$.

\section{Benchmarks}\label{sec:examples}

\begin{table*}[t]
\caption{A summary of numerical results for the benchmarks in Sec.~\ref{sec:examples}.
BH $n$-to-$k$ denotes Bravyi--Haah distillation of $k$ magic states from $n$ noisy inputs.
Color and RP$^2$ denote magic state cultivation on the color code and the RP$^2$ surface code,
and $d$ is the code distance.
3DCC to 2DCC and 3DCC to 2DSC denote code switching from the three-dimensional color code
to the two-dimensional color and surface codes.
TQD $L$, $T$ denotes the twisted-quantum-double protocol on an $L\times L$ periodic lattice with $T$ syndrome rounds.
Examples marked with an asterisk are not expanded in this paper; see Ref.~\cite{magna} for details.
The numbers of physical qubits and quantum gates are denoted by $N_q$ and $N_g$, respectively.
Each one-, two- or three-qubit gate application counts once.
Classically conditioned gates are counted by the quantum operation they apply.
Preparation, measurement, noise channels, classical processing and ideal final-round diagnostics are excluded.
The non-Clifford gate column $N_{\text{nC}}$ lists the counts and types of $T$, $\mathrm{C}S$ and $\mathrm{CC}Z$ gates,
counting $T^\dagger$ as $T$ and $\mathrm{C}S^\dagger$ as $\mathrm{C}S$.
The output column specifies the state, with tensor powers indicating multiple copies.
Encoded output states carry code subscripts: 2DCC, 2DSC and RP$^2$ as above, and TC for the toric code.
The representation ranks of the fidelity, acceptance and sampling networks
are denoted by $r_{\text{fid}}$, $r_{\text{acc}}$ and $r_{\text{samp}}$, respectively.
A dash indicates that no sampling network is needed.
Preparation time includes network construction and is incurred once before repeated sampling.
Times are medians over three runs, or ten for BH with $k=40$ and $60$,
each of $2048$ draws for BH, $1024$ for cultivation and code switching, and $256$ for TQD after one excluded warmup draw;
logical-level distillation is timed on its single batch of $16\,384$ draws.
For nonlinear feedback, sampling time is shown as
$T_{\text{samp}}(T_{\text{samp}}-T_{\text{dec}})$, where $T_{\text{dec}}$ is the classical decoding time,
measured over the same window and subtracted in each run before the median is taken.
Both values are per attempted draw.
}
\label{tab:examples-overview}
\begin{ruledtabular}
\footnotesize
\setlength{\tabcolsep}{2pt}
\begin{tabular}{lrrrlrrrrr}
Example & $N_q$ & $N_g$ & $N_{\text{nC}}$ & Output & $r_{\text{fid}}$ & $r_{\text{acc}}$ & $r_{\text{samp}}$ & Prepare (\unit{\milli\second}) & $T_{\text{samp}}$ (\unit{\milli\second}/draw) \\
\hline
BH 14-to-2 & 30 & 206 & $14T$ & $\ket{T}^{\otimes 2}$ & 1 & 1 & -- & 3.66 & 0.0069 \\
BH 20-to-4 & 44 & 355 & $20T$ & $\ket{T}^{\otimes 4}$ & 1 & 1 & -- & 5.95 & 0.0077 \\
BH 26-to-6 & 58 & 467 & $26T$ & $\ket{T}^{\otimes 6}$ & 1 & 1 & -- & 8.07 & 0.00865 \\
BH 38-to-10 & 86 & 812 & $38T$ & $\ket{T}^{\otimes 10}$ & 1 & 1 & -- & 13.1 & 0.0101 \\
BH 68-to-20 & 156 & 1702 & $68T$ & $\ket{T}^{\otimes 20}$ & 1 & 1 & -- & 27.8 & 0.014 \\
BH 128-to-40 & 296 & 3332 & $128T$ & $\ket{T}^{\otimes 40}$ & 1 & 1 & -- & 62.1 & 0.0223 \\
BH 188-to-60 & 436 & 5447 & $188T$ & $\ket{T}^{\otimes 60}$ & 1 & 1 & -- & 100 & 0.0312 \\
BH 248-to-80 & 576 & 7565 & $248T$ & $\ket{T}^{\otimes 80}$ & 1 & 1 & -- & 146 & 0.0412 \\
BH 308-to-100 & 716 & 10130 & $308T$ & $\ket{T}^{\otimes 100}$ & 1 & 1 & -- & 201 & 0.0503 \\
Color $d=3$ & 15 & 96 & $15T$ & $\ket{\bar{T}}_{\text{2DCC}}$ & 1 & 1 & -- & 2.62 & 0.00589 \\
RP$^2$ $d=3$ & 25 & 180 & $31T$ & $\ket{\bar{T}}_{\text{RP}^2}$ & 1 & 1 & -- & 3.75 & 0.00639 \\
Color $d=5$ & 42 & 547 & $53T$ & $\ket{\bar{T}}_{\text{2DCC}}$ & 1 & 1 & -- & 11.2 & 0.0104 \\
RP$^2$ $d=5$ & 61 & 640 & $101T$ & $\ket{\bar{T}}_{\text{RP}^2}$ & 1 & 1 & -- & 12.4 & 0.00877 \\
Logical-level distillation ($d=3$) & 102 & 2419 & $15T$ & $\ket{\bar{T}}_{\text{2DSC}}$ & 1 & 1 & 1 & 319 & 1.27 (0.64) \\
3DCC to 2DCC$^*$ & 38 & 149 & $15T$ & $\ket{\bar{T}}_{\text{2DCC}}$ & 1 & 1 & 1 & 4.37 & 0.0148 (0.0144) \\
3DCC to 2DSC$^*$ & 70 & 431 & $15T$ & $\ket{\bar{T}}_{\text{2DSC}}$ & 1 & 1 & -- & 10.7 & 0.00711 \\
TQD $L=3$, $T=5$$^*$ & 55 & 2032 & $288\mathrm{CC}Z$ & $\ket{\overline\psi_+}_{\text{TC}}$ & 4 & 2 & 1 & 438 & 3.2 (1.74) \\
TQD $L=4$, $T=6$$^*$ & 97 & 4420 & $608\mathrm{CC}Z$ & $\ket{\overline\psi_+}_{\text{TC}}$ & 4 & 2 & 1 & 1330 & 6.67 (4.43) \\
TQD $L=5$, $T=8$$^*$ & 151 &9038 & $1250\mathrm{CC}Z$ & $\ket{\overline\psi_+}_{\text{TC}}$ & 4 & 2 & 1 & 4550 & 16.5 (12.6) \\
\end{tabular}
\end{ruledtabular}
\end{table*}

In this section, we benchmark \simname\ with several examples.
Table~\ref{tab:examples-overview} summarizes the circuit sizes,
the input and output non-Clifford resources, the representation ranks of the reference networks,
and the measured run times for preparation and sampling.
We discuss three benchmarks in detail.
Bravyi--Haah distillation (Sec.~\ref{sec:bh-distillation}) scales up the number of output magic states.
Magic state cultivation (Sec.~\ref{sec:cultivation}) fixes the output to one logical magic state
while the circuit grows, and allows a comparison with other simulators.
Logical-level distillation (Sec.~\ref{sec:logical-distillation}) validates the sampling with complex nonlinear feedforward against an exact theoretical prediction.
Table \ref{tab:examples-overview} also contains more examples demonstrating the versatility of our method:
two code switching protocols \cite{daguerreCodeSwitchingRevisited2025, wangNonCliffordQuantumGate2026}
and a protocol based on the twisted quantum double (TQD) \cite{davydovaUniversalFaultTolerant2025, magdalenadelafuenteHighthresholdDecodingNonPauli2026},
a non-Pauli stabilizer code whose syndrome extraction contains $\mathrm{CC}Z$ gates
and decoder-controlled $X$ corrections in every round \footnote{The TQD output in Table~\ref{tab:examples-overview} is the four-logical-qubit state
$\ket{\overline\psi_+}\propto(I+\overline C)\ket{\overline+}^{\otimes4}$ on two toric codes,
where $\overline C$ is the product of two logical $\mathrm{C}Z$ gates between the codes
that the protocol measures, here with outcome $+1$.}.
All timing measurements were performed on an Apple Mac Studio
equipped with an M1 Ultra processor and \qty{128}{\gibi\byte} of unified memory.
Noise models, sample sizes, and importance-sampling proposals are reported in figure captions.
See Data Availability for more details.

\subsection{High-throughput magic state distillation}\label{sec:bh-distillation}

\begin{figure*}[tp]
\centering
\includegraphics{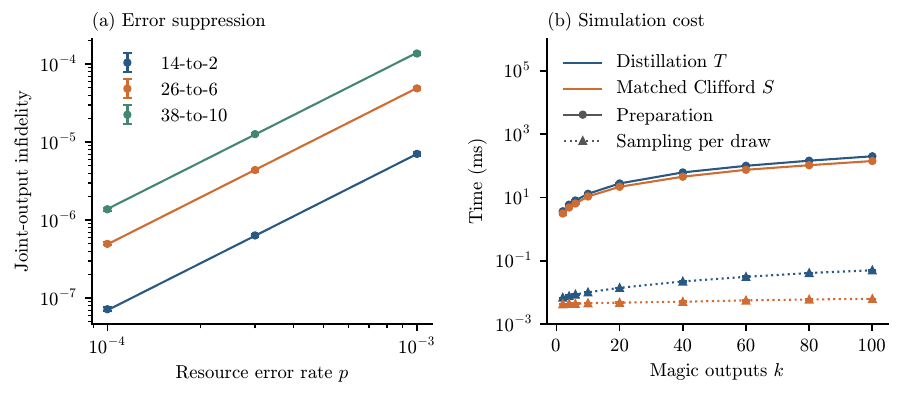}
\caption{High-throughput magic state distillation.
(a) Joint infidelity of the $k$ outputs for $k=2,6,10$,
with independent $Z$ errors on the input magic states and otherwise ideal operations.
Lines show the exact theoretical predictions, and markers show our Monte Carlo estimates
with pointwise $1.96$-standard-error intervals.
Each marker is an independent batch of $65\,536$ importance-sampled draws from a proposal
that raises the error probability of every input from $p$ to $2/(3k+8)$.
(b) Preparation time (circles) and sampling time per draw (triangles) at $p=10^{-3}$ up to $k=100$,
for the distillation circuit and for a matched Clifford circuit.
Markers show medians over repeated runs, with bars spanning their minimum-to-maximum ranges.
}
\label{fig:bh-distillation}
\end{figure*}

Magic state distillation has been the standard technique for fault-tolerant magic state preparation
since its invention \cite{bravyiUniversalQuantumComputation2005}.
Compared with recent proposals that prepare one logical $T$ state per run,
a key advantage is that a single distillation circuit can output many $T$ states.
Here we benchmark \simname's support for multi-magic-state output with the Bravyi--Haah
$(3k+8)$-to-$k$ family with even $k$~\cite{bravyiMagicStateDistillation2012},
up to $k = 100$.
In our simulation, all qubits are bare physical qubits that are not encoded in an underlying code.
Each input magic state has a $Z$ error with probability $p$,
and all other operations are noiseless.

We first validate the simulation.
Figure~\ref{fig:bh-distillation}(a) shows the simulated logical error rate,
defined as the infidelity with the joint target $\ket{T}^{\otimes k}$, for $k = 2,6,10$
and $p$ ranging from $10^{-4}$ to $10^{-3}$.
The data are obtained from $65\,536$ importance-sampled draws per circuit and error rate.
The estimates agree with the exact theoretical values
calculated from Eqs.~(14)--(16) of Ref.~\cite{bravyiMagicStateDistillation2012}
within $1.2$ standard errors, with relative standard errors between $2.0\%$ and $2.8\%$.

We then simulate $k = 2,4,6,10,20,40,60,80,100$ at $p = 10^{-3}$
to demonstrate the simulation of high-throughput protocols.
The circuit with $k$ output magic states consumes $3k+8$ input magic states,
so the largest instance contains 308 $T$ gates on 716 qubits.
Across all these circuits, the representation ranks of both reference networks remain 1.
Figure~\ref{fig:bh-distillation}(b) shows the preparation time and the sampling time per draw.
As a matched Clifford comparison, we replace the injected resources by $\ket S$
and use the corresponding Clifford injection corrections and a product-$S$ target,
with the same circuit width, noise model, and sampling settings.
The sampling time per draw grows from \qty{6.9}{\micro\second} at $k=2$ to \qty{50}{\micro\second} at $k=100$
(Table~\ref{tab:examples-overview}),
whereas that of the Clifford circuit stays between \qty{4.3}{\micro\second} and \qty{6.3}{\micro\second}.
The preparation time grows from \qty{3.7}{\milli\second} to \qty{201}{\milli\second}
and stays within a factor of $1.5$ of that of the Clifford circuit.
On the logarithmic time axis, both curves bend downward as $k$ increases,
so neither cost grows exponentially in $k$ over the tested range.

\subsection{Magic state cultivation}\label{sec:cultivation}

\begin{figure*}[tp]
\centering
\includegraphics{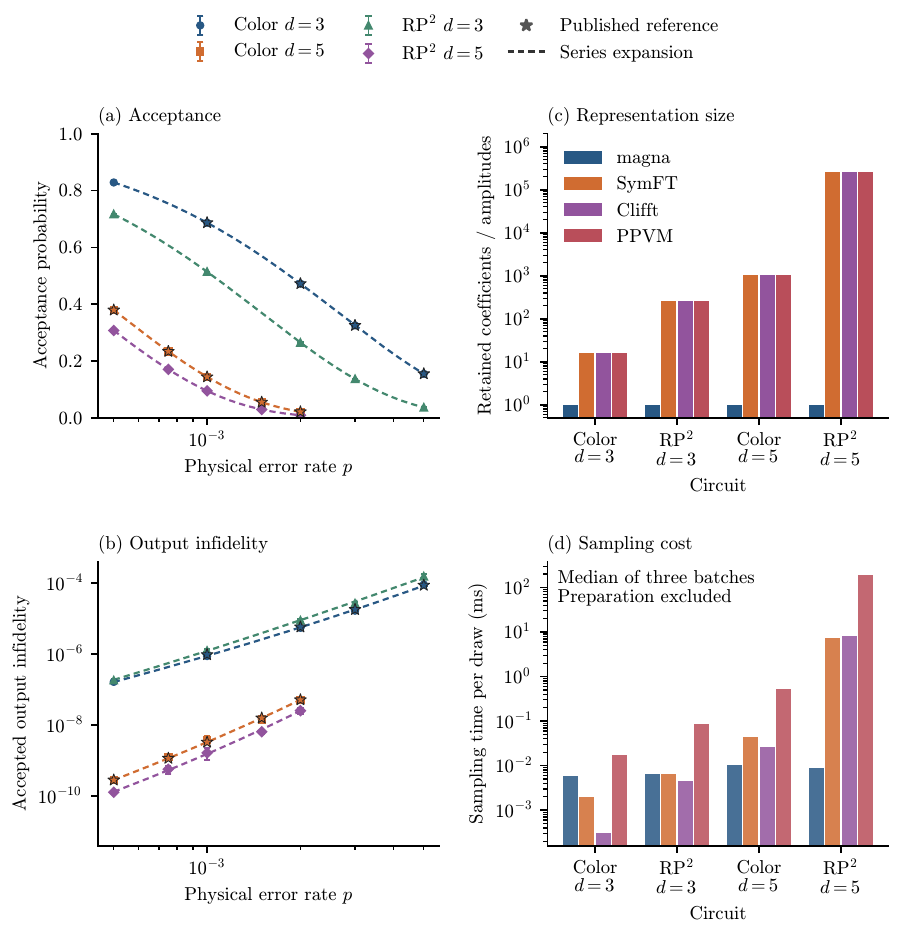}
\caption{Magic state cultivation on the color code (Color) and the RP$^2$ surface code at $d=3,5$.
(a) Acceptance rate and (b) output infidelity of accepted runs,
under the uniform depolarizing noise model of strength $p$ of Ref.~\cite{gidneyMagicStateCultivation2024}.
Stars are published values:
Ref.~\cite{gidneyMagicStateCultivation2024} for Color $d=3$
and Ref.~\cite{wanExactLogicalError2026} for Color $d=5$.
Dashed curves show the series expansions of Ref.~\cite{wanExactLogicalError2026} and its accompanying code.
Bars are pointwise $1.96$-standard-error intervals.
At each rate, the acceptance rate is the mean over $2^{24}$ draws from the physical fault distribution.
The accepted-failure probability is estimated from an independent batch
of $2^{22}$ draws at $d=3$ and $2^{27}$ at $d=5$,
by importance sampling from a proposal that raises the probability of each of the $N$ fault locations from $p$ to $p+3/N$
and, with probability $0.95$, chooses the last fault so that the configuration passes the detectors of the $S$ proxy.
For the RP$^2$ $T$ circuit at $d=5$, each batch has $2^{28}$ draws,
the proposal chooses the last two faults in this way,
and the batches contain $8976$--$15\,045$ accepted failures.
(c) Size of the representation stored by each simulator:
representation ranks of the reference networks for \simname,
peak number of amplitudes for SymFT and Clifft, and peak number of nonzero coefficients for PPVM.
(d) Sampling time per draw at $p=10^{-3}$, excluding preparation and warmup, as the median of three batches.
All simulators are given the same circuits, which are available in Ref.~\cite{magna}.}
\label{fig:cultivation}
\end{figure*}

\begin{table*}[t]
\caption{Simulation resources and sampling times for magic state cultivation at $p=10^{-3}$.
The circuit sizes $N_q$, $N_g$ and the non-Clifford gate counts $N_{\text{nC}}$
follow the conventions of Table~\ref{tab:examples-overview}.
For resource indicators, $R$ is the maximum representation rank of our reference networks, as defined in Sec.~\ref{sec:global-reduction},
$A_{\text{S}}$ and $A_{\text{C}}$ are the peak numbers of amplitudes stored by SymFT and Clifft,
and $A_{\text{P}}$ is the peak number of nonzero coefficients in PPVM.
Times are medians over three batches of $1024$ draws ($128$ for PPVM on RP$^2$ $d=5$).
They include noise generation and probability evaluation, and for PPVM its per-draw tableau construction,
and exclude compilation, planning, cache construction, and warmup.
We use SymFT at revision \texttt{c89b9851} of its public repository, Clifft 0.11.0,
and the generalized-tableau backend of PPVM at revision \texttt{7b2e1bcb} with coefficient threshold $10^{-10}$,
each through its published sampling interface with one thread,
and run them one at a time with a memory limit of \qty{64}{\gibi\byte}.}
\label{tab:cultivation-resources}
\begin{ruledtabular}
\small
\setlength{\tabcolsep}{2pt}
\begin{tabular}{lrrrrrrrrrrr}
\multicolumn{8}{c}{} & \multicolumn{4}{c}{Sampling time (\unit{\milli\second}/draw)} \\
Circuit & $N_q$ & $N_g$ & $N_{\text{nC}}$ & $R$ & $A_{\text{S}}$ & $A_{\text{C}}$ & $A_{\text{P}}$ & \simname & SymFT & Clifft & PPVM \\
\hline
Color $d=3$ & 15 & 96 & $15T$ & 1 & 16 & 16 & 16 & 0.00589 & 0.00202 & 0.000309 & 0.0173 \\
RP$^2$ $d=3$ & 25 & 180 & $31T$ & 1 & 256 & 256 & 256 & 0.00639 & 0.00655 & 0.00456 & 0.0875 \\
Color $d=5$ & 42 & 547 & $53T$ & 1 & 1024 & 1024 & 1024 & 0.0104 & 0.0436 & 0.0267 & 0.531 \\
RP$^2$ $d=5$ & 61 & 640 & $101T$ & 1 & 262144 & 262144 & 262144 & 0.00877 & 7.46 & 8.14 & 193 \\

\end{tabular}
\end{ruledtabular}
\end{table*}

Magic state cultivation \cite{gidneyMagicStateCultivation2024} is an efficient protocol
for preparing a magic state in a finite-size surface code.
Here we consider two protocols:
the original protocol, based on measuring a logical Clifford gate on a color code
\cite{gidneyMagicStateCultivation2024},
and a variant based on measuring a logical Clifford gate on a surface code
on the projective plane RP$^2$, which we call the RP$^2$ code
\cite{chenEfficientMagicState2025}.
We simulate both protocols at $d=3$ and $d=5$.
The number of $T$ gates in the circuit grows with the code distance $d$,
from 15 to 53 for the color code and from 31 to 101 for the RP$^2$ code.
Cultivation therefore tests directly how the simulation resource follows the input non-Clifford resources for circuits with similar structure.
It has also been simulated with several recent simulators,
which allows a comparison on identical circuits.

For color-code cultivation,
we use the circuit of Ref.~\cite{gidneyMagicStateCultivation2024} with unitary injection for $d=3$,
and that of Ref.~\cite{wanExactLogicalError2026} for $d=5$.
The RP$^2$ circuits are taken from Ref.~\cite{chenEfficientMagicState2025}.
Both protocols continue with an escape stage that transfers the state to a larger code.
We do not simulate this stage,
and neither do the published results that we compare with.
The circuits include the feedforward and postselection of their sources
and their uniform depolarizing noise model of strength $p$ \cite{gidneyMagicStateCultivation2024}.
For all circuits, we report the infidelity of the accepted logical qubit with $\ket{T}$.
The color-code source circuits end with a noiseless logical measurement in the basis of the target state,
whose error probability, reported in the published results, equals this infidelity.

We first validate the simulation against published results.
Figures~\ref{fig:cultivation}(a) and~\ref{fig:cultivation}(b) show
the acceptance rate and the output infidelity of accepted runs.
For all points in Figs.~\ref{fig:cultivation}(a) and~\ref{fig:cultivation}(b), the acceptance is sampled directly,
and the infidelity is obtained from an independent estimate of the accepted-failure probability.
For the color code, the published values are the state-vector samples of
Ref.~\cite{gidneyMagicStateCultivation2024} at $d=3$
and the exact values of Ref.~\cite{wanExactLogicalError2026} at $d=5$.
Our infidelity estimates differ from these values by at most $1.8$ combined standard errors,
and our acceptance estimates by at most $1.5$.
For the RP$^2$ code, we compare with the series expansions of Ref.~\cite{wanExactLogicalError2026}
for the same circuits with $T$ gates.
Our acceptance estimates agree with them within $1.3$ standard errors,
and our infidelity estimates within $2.3$ standard errors.

We next compare the simulation resource with that of other simulators.
We run SymFT \cite{fangSymFTUniversalFaultTolerant2026}, Clifft \cite{chaseClifftFastExact2026},
and PPVM~\cite{luoPPVMCircuitSimulation2026}
on the same four circuits.
Figure~\ref{fig:cultivation}(c) shows the size of the representation that each simulator stores.
For SymFT, Clifft, and PPVM, the peak sizes coincide,
and they grow from 16 to 1024 between $d=3$ and $d=5$ for the color code
and from 256 to $262\,144$ for the RP$^2$ code.
The cumulative numbers of dense rotation updates in SymFT grow
from 222 to $38\,982$ for the color code and from 6024 to $12\,746\,692$ for the RP$^2$ code.
We also tested Tsim~\cite{haenelTsimFastUniversal2026} with the cutting procedure of Ref.~\cite{wanSimulatingMagicState2025},
which completes the sampling only for Color $d=3$:
for RP$^2$ $d=3$, its warmup encounters 87 fault-pattern bits, beyond the 63-bit hashing limit of the implementation,
and the $d=5$ circuits have components with 15 and 31 outputs, beyond its default enumeration limit of 12.
In contrast, both of our reference networks have representation rank one for all four circuits.

The sampling times per draw at $p=10^{-3}$ in Fig.~\ref{fig:cultivation}(d) and Table~\ref{tab:cultivation-resources}
follow these sizes.
Ours stays between \qty{5.9}{\micro\second} and \qty{10.4}{\micro\second} for all four circuits.
At $d=3$, where SymFT and Clifft store at most 256 amplitudes, they are as fast as our method or faster.
At $d=5$, our method is faster than every other simulator,
by about three orders of magnitude on the RP$^2$ code, where SymFT and Clifft store $2^{18}$ amplitudes.
Here we compare the time per draw rather than the time to reach a given precision of the logical error rate.
The comparison is conservative for our method,
because for each fault configuration it returns the acceptance and failure probabilities
averaged over all measurement outcomes, rather than one sampled measurement record,
which removes the variance due to the outcomes.

\subsection{Logical-level distillation}\label{sec:logical-distillation}

In practice, magic state distillation is carried out on logical qubits,
so its feedforward and postselection are conditioned on decoded outcomes.
We test the support of \simname\ for such nonlinear rules
on a surface-code implementation of 15-to-1 distillation
\cite{fowlerSurfaceCodesPractical2012,litinskiGameSurfaceCodes2019},
which contains fifteen serial rounds of nonlinear feedforward and a final nonlinear postselection,
under an error model whose logical error rate can be calculated exactly.

The circuit uses five distance-three working patches and one resource patch, on 102 physical qubits,
in the space-time-efficient form of 15-to-1 distillation \cite{haahCodesProtocolsDistilling2018,litinskiGameSurfaceCodes2019}.
It consumes fifteen encoded $T$ states one after another,
each prepared on the resource patch, which is reset and reused.
Each $T$ state is coupled to the working patches by transversal CNOT gates and then measured.
A decoder obtains the logical outcome from the physical measurement outcomes,
and this outcome controls a fold-transversal $S^\dagger$ gate on the working patches
\cite{moussaTransversalCliffordGates2016,chenTransversalLogicalClifford2026}
before the next $T$ state is consumed.
The output is one encoded $T$ state,
which is accepted when the four checks of the distillation code pass.
In the error model, each input $T$ state is encoded without error,
every data qubit of its patch is then depolarized with total probability $p$,
and one noiseless round of syndrome extraction and matching recovery follows.
All other operations are ideal.
With this error model,
each input then carries an effective logical phase error with probability
$q_{\text{eff}}=P_Z+(P_X+P_Y)/2$,
where $P_X$, $P_Y$, and $P_Z$ are the exact probabilities of the residual logical errors of one input,
obtained by enumerating all Pauli patterns on its patch,
and a logical $X$ or $Y$ error acts as a phase error with probability one half in the injection.
With $s=1-2q_{\text{eff}}$, the exact acceptance rate and infidelity are 
\cite{bravyiUniversalQuantumComputation2005, haahCodesProtocolsDistilling2018}:
\begin{equation}
 p(\text{acc})=\frac{1+15s^8}{16},\qquad
 1-F=\frac{1+15s^8-15s^7-s^{15}}{2(1+15s^8)},
 \label{eq:logical-distillation-exact}
\end{equation}
with which we compare the simulation of the complete physical circuit.

\begin{figure*}[tp]
\centering
\includegraphics{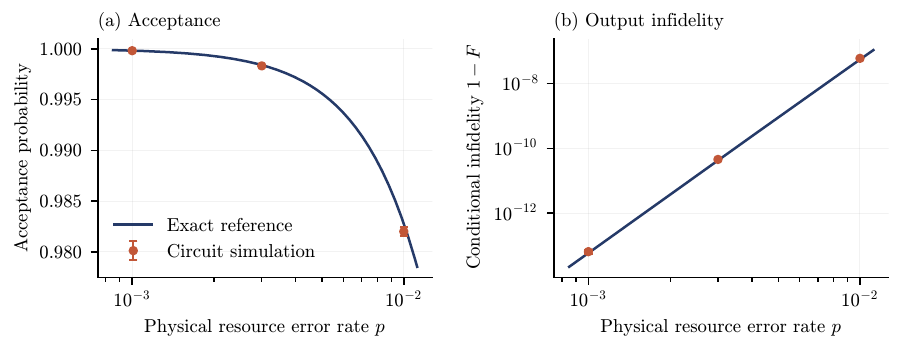}
\caption{Logical-level 15-to-1 distillation on six distance-three surface-code patches,
with fifteen stages of decoder-controlled feedforward and a final decoder-controlled postselection.
(a) Acceptance rate and (b) infidelity of the accepted encoded output,
as functions of the depolarizing probability $p$ on the data qubits of each encoded input $T$ state.
Curves are the exact predictions of Eq.~\eqref{eq:logical-distillation-exact}.
Markers show the simulation of the full physical circuit with $16\,384$ importance-sampled draws with one-standard-error bars.
For each patch, the proposal assigns probability $0.2$ to a nonidentity residual logical class,
split among $X$, $Y$, and $Z$ in their physical proportions at $p=10^{-2}$,
and draws the physical Pauli pattern from its exact conditional distribution;
the decoders receive only the reported syndromes.
All sampled measurement histories have weight one, as expected because the sampling network has $d_W=0$ (Sec.~\ref{sec:nonlinear-feedforward}).}
\label{fig:logical-distillation}
\end{figure*}

Figure~\ref{fig:logical-distillation} compares the simulation with the exact curves
at $p=10^{-3}$, $3\times10^{-3}$, and $10^{-2}$.
A single batch of $16\,384$ importance-sampled draws is reweighted to all three error rates.
At every rate, the estimated infidelity agrees with the exact value within $1.0$ standard error,
including $p=10^{-3}$, where the infidelity is only $6\times10^{-14}$.
All three reference networks reduce to one stabilizer term.
Preparation takes \qty{319}{\milli\second}, and sampling takes \qty{1.27}{\milli\second} per draw,
of which about half is spent in the decoder.

\section{Summary and discussion}\label{sec:discussion}

We have introduced \simname, a classical simulator that evaluates magic state factories
with their full non-Clifford circuits at a cost comparable to that of Clifford circuits.
For Bravyi--Haah distillation, it simulates instances with up to 100 output magic states,
308 $T$ gates, and 716 qubits,
with a preparation time within a factor of $1.5$ of a matched Clifford circuit
and \qty{50}{\micro\second} per draw at the largest size.
For distance-5 magic state cultivation, it samples in about \qty{10}{\micro\second} per draw
on both the color code and the RP$^2$ code,
about three orders of magnitude faster than state-vector-based simulators on the latter.
It also treats nonlinear feedforward faithfully:
for a surface-code 15-to-1 distillation with fifteen rounds of decoder-controlled corrections,
it reproduces the exact logical error rate down to $6\times10^{-14}$.
In every protocol of Sec.~\ref{sec:examples}, most reference networks reduce to a single stabilizer term, and none requires more than four.

These results remove several obstacles that have constrained the study of magic state factories.
Protocols can be evaluated with their actual $T$ gates rather than with Clifford proxies,
which have been shown to be unreliable for magic state cultivation \cite{liSOFTHighperformanceSimulator2025, wanExactLogicalError2026, chanDiagnosingRestoringDegraded2026}.
High-throughput factories, beyond the reach of simulators that represent the output state,
can now be assessed at realistic sizes.
Protocols with decoder-dependent feedforward, essential for logical-level factories,
can be simulated without rewriting their classical control as quantum gates.
Because a single preparation serves all draws and each draw costs microseconds,
sweeps over noise rates, decoders, and postselection rules become routine.

More broadly, our results show that circuits with a large number of non-Clifford gates
can be simulated efficiently when those gates act as a single logical layer.
Existing classical simulators generally have costs governed by the non-Clifford content of a circuit,
quantified for instance by the stabilizer rank \cite{bravyiImprovedClassicalSimulation2016}
or by the negativity of quasiprobability representations \cite{pashayanEstimatingOutcomeProbabilities2015}, 
which has also been used to characterize the dynamics of magic in random circuits \cite{zhangQuantumMagicDynamics2024},
and the number of non-Clifford gates needed for Clifford circuits to approximate unitary designs
has been characterized \cite{zhangDesignsMagicaugmentedClifford2025}.
This raises a general question:
is there a measure of classical simulation hardness that captures
how the non-Clifford gates are organized within a circuit?  
On the technical side, extending the gate set to Pauli rotations by general angles
would allow coherent errors to be simulated.
Such errors may retain the single-layer logical structure,
although the phase analysis and its Pauli-covariance would need to be generalized.
The remaining non-Clifford coordinates could also be summed by tensor-network contraction~\cite{markovSimulatingQuantumComputation2008}
instead of being expanded into stabilizer terms,
combining a cost limited by entanglement with one limited by non-Clifford resource,
as explored for state-evolution simulators~\cite{hartwegComparingMagicState2026}.

Most importantly, we expect \simname\ to change how magic state factories are developed.
Until now, simulation has mostly served to verify a protocol after it was designed,
often with Clifford proxies or at small sizes,
because faithful simulation of the full non-Clifford circuit was too costly to repeat.
With per-draw costs comparable to Clifford simulation,
the full circuit can instead sit inside a design loop,
as fast analytical fault-distance checks already do for cultivation~\cite{chanDiagnosingRestoringDegraded2026}:
candidate protocols are generated, simulated with their actual $T$ gates under realistic noise,
and refined according to the measured acceptance rate and fidelity.
Code choices, syndrome-extraction schedules, postselection rules, and decoders
can then be optimized jointly,
by systematic search or by automated optimization,
and tailored to the noise models of specific hardware platforms.
Such a loop can explore far more designs than analytical estimates allow,
and we expect it to accelerate the search for factories with lower space-time overhead,
one of the central costs of fault-tolerant quantum computation.

\section*{Data availability}

The source code and user manual of \simname,
together with the circuit files of all examples in Table~\ref{tab:examples-overview},
are available in Ref.~\cite{magna}.
The random seeds, raw sampling data, and figure scripts are available in Ref.~\cite{magnaData}.

\section*{Acknowledgment}
We thank Hong-Ye Hu, Kwok Ho Wan, Ainhoa Zapirain, Zikang Zhang and Tian-Gang Zhou for useful discussions. This work is supported by the National Natural Science Foundation of China under Grant No.~12575022, Scientific Research Innovation Capability Support Project for Young Faculty under Grant No.~ZY2025157, the Shanghai Committee of Science and Technology under Grant No.~25LZ2600800, and the Tsinghua Dushi Program. 

\emph{Note added.---} Near the completion of this work, we became aware of
several related works on the simulation of noisy magic state preparation
\cite{takadaExactEfficientSimulation2026,xiaoEfficientFidelitySimulation2026,cercelescuPolynomialtimeSimulationNonClifford2026,luoPPVMCircuitSimulation2026}.
Like our method, \texttt{merlin}~\cite{cercelescuPolynomialtimeSimulationNonClifford2026}
and the method of Ref.~\cite{xiaoEfficientFidelitySimulation2026}
avoid an exponential cost in the number of output magic states,
but their efficiency requires restricted gates and feedforward,
such as Pauli, CNOT and $T$ gates with diagonal Clifford corrections~\cite{cercelescuPolynomialtimeSimulationNonClifford2026}
or $X$, CNOT and diagonal gates~\cite{xiaoEfficientFidelitySimulation2026}.
PPVM~\cite{luoPPVMCircuitSimulation2026}, like our method,
supports feedforward conditioned on decoded outcomes,
but only in its generalized-tableau backend,
whose cost can grow exponentially with the number of non-Clifford gates.
Our method avoids both limitations:
it accepts arbitrary Clifford gates, $T$, $\mathrm{C}S$ and $\mathrm{CC}Z$ gates,
and Clifford feedforward and postselection controlled by arbitrary classical rules,
while its simulation cost remains low for all benchmarks in Table~\ref{tab:examples-overview},
which span the main families of magic state factories (distillation, cultivation and code switching)
and include a protocol on a non-Pauli stabilizer code.

\bibliography{references}
\end{document}